\documentclass[aps,pre,twocolumn,groupedaddress,showpacs]{revtex4-1}

\usepackage[utf8]{inputenc}
\usepackage[english]{babel}
\usepackage{amsmath,graphicx,enumerate}
\begin{document}

\title{Linear and non-linear thermodynamics of a kinetic heat engine with fast transformations.} 
\author{Luca Cerino}
\author{Andrea Puglisi}
\author{Angelo Vulpiani}
\affiliation{Istituto dei Sistemi Complessi - CNR and Dipartimento di Fisica, Universit\`a di Roma Sapienza, P.le Aldo Moro 2, 00185, Rome, Italy}

\date{\today}

\begin{abstract}
  We investigate a kinetic heat engine model constituted by particles
  enclosed in a box where one side acts as a thermostat and the
  opposite side is a piston exerting a given pressure. Pressure and
  temperature are varied in a cyclical protocol of period $\tau$:
  their relative excursions, $\delta$ and $\epsilon$ respectively,
  constitute the thermodynamic forces dragging the system
  out-of-equilibrium. The analysis of the entropy production of the
  system allows to define the conjugated fluxes, which are
  proportional to the extracted work and the consumed heat. In the
  limit of small $\delta$ and $\epsilon$ the fluxes are linear in the
  forces through a $\tau$-dependent Onsager matrix whose off-diagonal
  elements satisfy a reciprocal relation. The dynamics of the piston
  can be approximated, through a coarse-graining procedure, by a
  Klein-Kramers equation which - in the linear regime - yields
  analytic expressions for the Onsager coefficients and the entropy
  production. A study of the efficiency at maximum power shows that
  the Curzon-Ahlborn formula is always an upper limit which is
  approached at increasing values of the thermodynamic forces,
  i.e. outside of the linear regime. In all our analysis the adiabatic
  limit $\tau \to \infty$ and the the small force limit
  $\delta,\epsilon \to 0$ are not directly related.
\end{abstract}

\pacs{05.70.Ln,05.40.-a,05.20.-y}


\maketitle
\section{Introduction}

Thermodynamics, at its origins, received a crucial impulse from the
study of heat engines~\cite{carnot}. It is interesting to realize that
- after almost two centuries - engines still represent a relevant
driving force towards new developments in this science. A challenging
frontier in thermodynamics is the world of small and fast systems,
where the assumptions of ``quasi-reversible'' transformations and the
thermodynamic limit of statistical mechanics are not
valid~\cite{VandenBroeck2005,Schmiedl2008,cerino15}. Obviously,
``fast'' thermodynamic transformations, i.e. those such that the
typical cycle time $\tau$ is shorter than the slowest relaxation time
of the system, constitute a key problem in the industry and, for this
reason, have been under the scrutiny for many decades. In the 70's of
the last century, several results were obtained in the so-called
finite time thermodynamics~\cite{Andresen1977}, one of the foremost
being the Curzon-Ahlborn (CA) estimate for the efficiency at maximum
power~\cite{curzon1975efficiency}. Such an estimate has been revised
in the recent years, with the introduction of new and more general
classes of engines with respect to the original model considered by
Curzon and
Ahlborn~\cite{VandenBroeck2005,Esposito2009,Esposito2010}. A more
recent wave of studies concerns the thermodynamics of systems with a
``small'' number $N$ of degrees of freedom~\cite{klages}, motivated by
the tremendous increase of resolution in the observation and in the
manipulation of the micro-nano world, involving mainly biophysical
systems and artificial machines~\cite{Gaspard2006}. The distinguishing
feature of small systems is the relevance of fluctuations, which are
negligible only when the number of constituents is very large, as for
macroscopic bodies. The study of fluctuations in thermodynamics
functions such as energy or entropy goes back to Einstein, Onsager and
Kubo, but has recently taken an acceleration with the establishing of
new results in response theory~\cite{marconi2008fluctuation} and in
the so-called stochastic
thermodynamics~\cite{Seifert,Sekimoto2010}. Such a turning point
concerns the properties of fluctuations in system which are {\em far}
from thermodynamic equilibrium, and therefore is intimately connected
to the problem, mentioned before, of fast transformations. In a
nutshell, modern stochastic thermodynamics addresses the finiteness of
both transformation's time $\tau$ and system's size $N$.

In the literature about stochastic thermodynamics a prominent role is
covered by models, often inspired by minimal experiments at the
microscale, with very few degrees of freedom, where typically one has
$N = 1$: the overdamped dynamics of the position of a colloid in a
non-conservative (e.g. time-dependent) potential is a seminal
prototype~\cite{Seifert2005}. Only a few studies have discussed the
non-trivial effects of inertia~\cite{memory,Celani2012,bauer16} where the
relevant degrees of freedom are at least two (also with different
parities under time-reversal). It is even more rare to find models
with $N \gg 1$, still remaining in the domain of small $N$: for
instance with an order of magnitude $N \sim 10^2$ fluctuations can
still be relevant and possibly non-trivial, while the complexity of
the dynamics is hugely raised. Such numbers are also closer to real
biophysical applications with macromolecules, nanocapillaries,
etc.~\cite{cecconi}. On the front of the statistical mechanics of
molecular models, an exception is certainly represented by the study
in~\cite{Izumida2008,Izumida2009}, and by our more recent proposal
in~\cite{cerino15}. These papers investigate the dynamics of a gas
model with $N$ particles enclosed between a thermostat and a piston:
the piston is controlled through a cyclic protocol of duration $\tau$
that defines operations similar to a heat engine. The basic
equilibrium properties (i.e. thermodynamics and fluctuations, when
$\tau \to \infty$) of that particular gas-piston system have been
detailed in~\cite{Tfluct,cerino2014fluctuations}. In the study of the
cyclic protocols the papers~\cite{Izumida2008,Izumida2009} do not take
into account the piston's inertia and directly fix its
(time-dependent) position during the cycle. In~\cite{cerino15}, on the
contrary, we only fix the (time dependent) force acting on the piston,
so that the piston's velocity is determined by the effect, mediated by
inertia, of such a force and of the collisions with the gas'
molecules. Because of a larger freedom in the piston's dynamics, this
``machine'' displays a much richer diagram of phases. In particular,
the choice of $\tau$ determines different working regimes: engine,
refrigerator and heat pump.  In addition, also in view of an analysis
of the linear regime of the engine similar to the one discussed in
\cite{Izumida2009}, in this model it is possible to disentangle
the smallness of the external perturbation (represented by the
excursion of forces and temperatures) with the slowness of the
transformations (represented by the total time of the cycle).

In the present paper we study aspects of the model introduced
in~\cite{cerino15} which were not discussed or deepened in that
paper. Among the new results, we introduce a formalization of the
model in terms of fluxes and thermodynamics forces which allows to
distinguish between a linear and a non-linear regime. In the linear
regime, we get a matrix of Onsager coefficients which non-trivially
depend on $\tau$, a fact usually ignored in the recent
literature. Along those lines, we can widen the study of the power
optimization, considering different procedures of maximization
(e.g. by varying different parameters, including the cycle duration
$\tau$) and comparing the results with the CA estimate for the
efficiency.

The organization of the paper is sketched in the following. In
Section~\ref{sec:model} we illustrate the kinetic model and its
coarse-grained approximations, which are useful in the rest of the
paper, giving a quick overview of the results published
in~\cite{cerino15}. In Section~\ref{linear} we introduce the
thermodynamic analysis of the model, introducing the thermodynamic
forces, the entropy production and the fluxes, focusing on the linear
regime in the limit of small forces.  In Section~\ref{sec:eff_max_pow} we
discuss the efficiency at maximum power, under different protocols of
maximization, and compare it with the Curzon-Ahlborn estimate. Finally
we draw our conclusions and discuss perspectives for future studies in
Section~\ref{concl}.

\section{The engine and its phase diagram}
\label{sec:model}

In~\cite{cerino15} we have studied a molecular model (MM) of
engine. In the same paper we have also described an approximation of
the MM called 2V model, indicating that two macro-variables are used
to describe the coarse-grained dynamics of the engine. In view of the
present study, it is useful to summarize the key results of that
study.

{\em MM} The full model is an ideal gas of $N$ point-like
particles in a container with mass $m$, position ${\bf x}_i$ and
momentum ${\bf p}_i$ (${\bf v}_i={\bf p}_i/m$), $i=1\ldots N$ . The
real dimensionality of the box is not relevant, as the particles
interact only with the piston and the thermostat: we consider only the
$\hat{x}$ direction, assuming that the thermostat is at position $0$
and the piston is at position $X(t)>0$.  The piston has mass $M$ and
momentum $P$ and moves along $\hat{x}$ under the influence of the
collisions with the gas particles and of an externally controlled
force $\vec{F}(t)=-F(t)\hat{x}$.  If
${\boldsymbol{\Gamma}=(x_1,\ldots, x_N,p_1,\ldots,p_N,X,P)}$ is the
full phase space variable of the system, the Hamiltonian reads
\begin{equation}\label{eq: hamiltonian}
\mathcal{H}(\boldsymbol{\Gamma},t)=\sum_{i=1}^N \frac{p_i(t)^2}{2m}+\frac{P(t)^2}{2M} +F(t)X(t),
\end{equation}
with the additional constraints $0<x_i(t)<X(t)$, $i=1,\ldots,N$, and $X(t)>0$.
The collisions with the piston are assumed to be elastic,
i.e. conserve momentum and kinetic energy (see~\cite{cerino15} for the
details). When a particle collides with the wall at $x=0$ it takes
a new velocity $v'$  with probability density
\begin{equation}\label{eq: thermostat}
\rho(v')=\frac{m}{T_o}v'e^{-\frac{m v'^{2}}{2T_o}}\Theta(v),
\end{equation}
where $\Theta(v)$ is the Heaviside step function and we are measuring
temperature $T$ in energy units, i.e. $k_B=1$ for the Boltzmann
factor. When force and temperature are constant,
the stochastic dynamics generated by this rule satisfies the detailed
balance condition with respect to the canonical probability distribution
$\rho(\boldsymbol{\Gamma})\propto\exp(-\beta \mathcal{H})$.

Even if the particles do not directly interact, there is an indirect
but relevant interaction through the piston, making $N$ and $Nm/M$
important parameters for the
dynamics~\cite{lieb99,lesne06,cencini07,sasa15}. We study the system
in a range of $N m/M$ close to $\sim 1$, meaning that there is a
non-trivial interplay between the gas and the piston. The static study
of the system (equilibrium at fixed $F$ and $T_0$) can be found
in~\cite{cerino2014fluctuations,Tfluct}, yielding for the piston
position average $\langle X \rangle^{eq} =(N+1)T_o/F$ and variance
$\sigma^2_X=(N+1)T_o^2/F^2$. The average instantaneous kinetic
temperature of the gas $T(t)=\frac{1}{N}\sum_{i=1}^N m v_i(t)^2$, has
ensemble average $\langle T\rangle^{eq}=T_o$ and ensemble variance
$\sigma^2_T=2\,T_o^2/N$.

\begin{figure}[!hbtp]
\includegraphics[width=0.6\columnwidth]{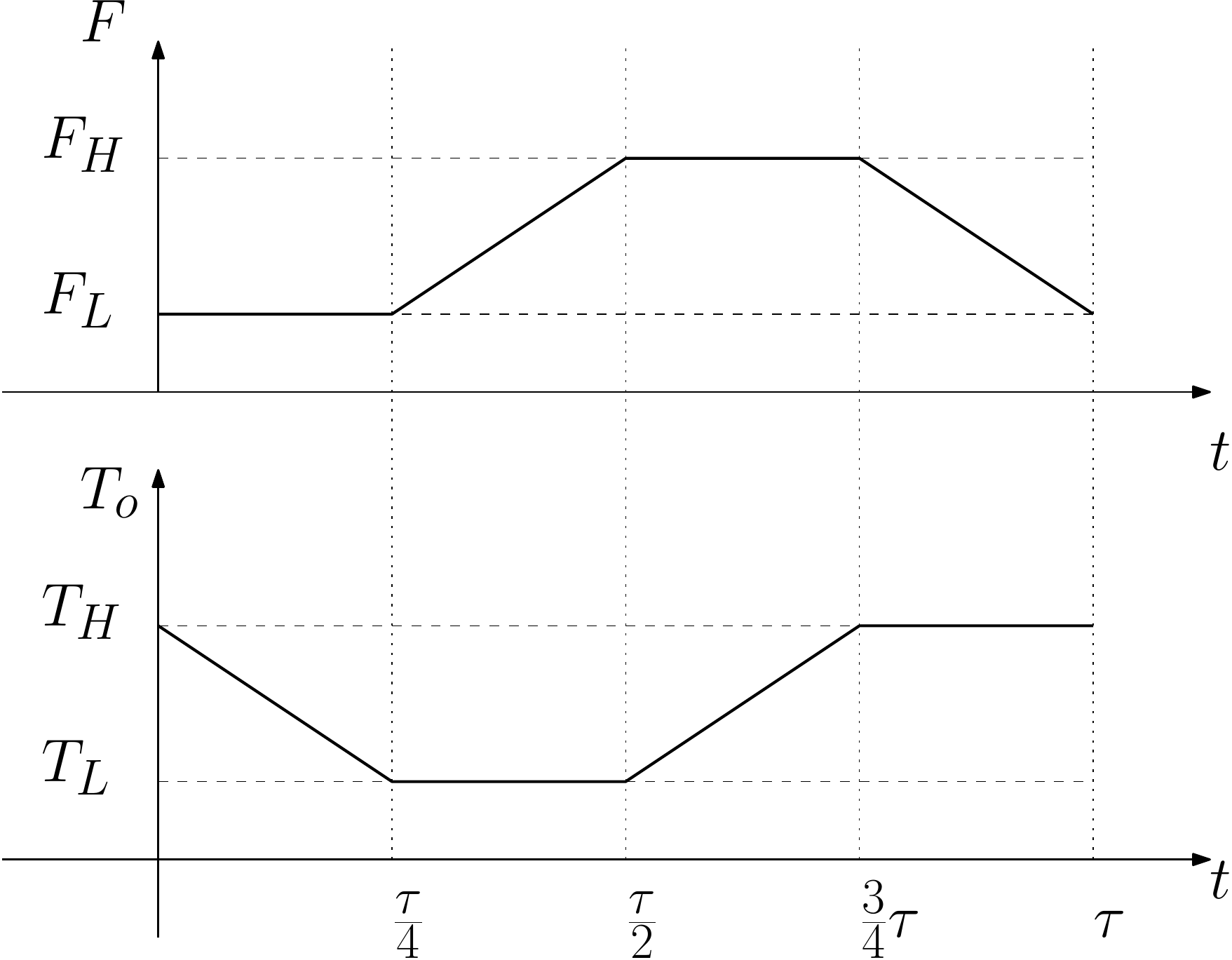}
\caption{Force $F$ and temperature $T_0$ as  functions of time over a cycle 
period $\tau$.}\label{fig:protocol}
\end{figure}
In order to obtain a heat engine (that is to extract mechanical work from
the thermostat), it is necessary to vary $F$ and $T_o$ with time. Here
we adopt the same cyclical protocol - also known as ``second type
Ericsson cycle'' - as in~\cite{cerino15}: the duration of a cycle,
$\tau$, is divided in $4$ sub-cycles, alternating two isobaric and two
isothermal transformations. Temperature and force variations are
always linear in time (see Fig. \ref{fig:protocol}).

The main reason for such a protocol is to have a system which is
always thermostatted, i.e. always near a canonical ensemble (at least
for slow transformations): on the contrary, adiabatic transformations
require a micro-canonical analysis which may become less transparent.
A similar model has been studied in~\cite{Izumida2008} with a crucial
difference: there the velocity of the piston is fixed at any time and
cannot fluctuate (this choice corresponds to the infinite mass $M$
limit). In~\cite{cerino15} we have shown that a finite mass of the
piston implies a rich phase diagram, due to the possibility of
oscillations out of phase with respect to the external force.

The thermodynamic variables associated to energy variations are easily
identified: the instantaneous internal energy
$E(t)=\mathcal{H}(\boldsymbol{\Gamma}(t),t)$, the input power
$\dot{W}(t)=\left.\frac{\partial \mathcal{H}(t)}{\partial
    t}\right|_{{\bf x}(t)}$
and the rate $\dot{Q}(t)$ of energy absorption from the thermal
wall. Conservation of energy implies
$\dot{E}(t)=\dot{Q}(t)+\dot{W} (t)$. For the Hamiltonian given in
Eq. (\ref{eq: hamiltonian}) one gets $\dot{W}=X\dot{F}$. Let us remark
that this formula is different from the one obtained in standard
thermodynamics, $\dot{W}=F\dot{X}$: this is due to the fact that we
included the energy of the piston in the internal energy of the system
\cite{Jarzynski2007}. For our choice of thermostat discussed before,
the formula for the energy adsorption (heat flux) can be formally
written as $\dot{Q}(t)=m\sum_i \delta(t-t_i) [(v_i'(t))^2-(v_i(t))^2]$
where $t_i$ are the times of collisions between the gas particles and
the thermostatting wall at $x=0$, whereas $v_i$ and $v_i'$ are the
velocities before and after a collision respectively.
\begin{table}[!tbp]
\centering
\begin{tabular}{c|c|c}
Segment&$\langle W\rangle$&$\langle Q\rangle$\\
\hline
I)&0&$\frac{3}{2}(N+1)(T_C-T_H)$\\
II)&$(N+1)T_C\ln\left(\frac{F_H}{F_L}\right)$&$-(N+1)T_C\ln\left(\frac{F_H}{F_L}\right)$\\
III)&0&$\frac{3}{2}(N+1)(T_H-T_C)$\\
IV)&$-(N+1)T_H\ln\left(\frac{F_H}{F_L}\right)$&$(N+1)T_H\ln\left(\frac{F_H}{F_L}\right)$\\
\end{tabular}
\caption{Table with the adiabatic values of $Q$ and $W$ in each 
segment of the Ericsson cycle. The average $\langle \cdot\rangle$ is 
intended over many realization of the cycle.}\label{tab: 1}
\end{table}
In the following, unless differently specified, we use the symbols $W$
and $Q$ to mean a time-integral over a cycle,
i.e. $W=\int_t^{t+\tau}\dot{W}(s)ds$ and
$Q=\int_t^{t+\tau} \dot{Q}(s)ds$. Due to the stochastic nature
of collisions and of the random choice of initial conditions, $W$ and
$Q$ are random variables. Conversely, the symbols $\langle W \rangle$
and $\langle Q \rangle$ indicate the average work and heat per cycle
computed over a long (single) run composed of a large number of
cycles.  Due to the periodic nature of the protocol the system will
reach, after a transient, a periodic asymptotic state with a
probability distribution in the full phase space depending on time
only through ${t'=t \mod \tau}$.  Therefore, thanks to the ergodic
hypotheses, the average denoted by $\langle \cdot \rangle$, is
equivalent to an average over the aforementioned periodic
distribution.

The study of the thermodynamics of the engine in the quasi-static
limit, i.e. assuming that the system is always at equilibrium:
$X(t)=\langle X \rangle^{eq}_{F(t),T_o(t)}$ and $E(t)=\langle
\mathcal{H}\rangle^{eq}_{F(t),T_o(t)}$, leads to the formula in Table
\ref{tab: 1}.  During segments I) and III) no work is done on the
system and the heats exchanged have same magnitude but opposite
signs. Therefore, in the quasi-static limit $\tau \to \infty$, there
is no net heat exchange with the intermediate reservoirs at
temperature $T_C<T^*<T_H$. We do not expect this ``heat symmetry'' to
hold when $\tau$ is finite. However, since the observed discrepancies
are not large, in \cite{cerino15} we identified the input heat
$Q_{in}$ with the energy absorbed from the hot reservoir at $T_H$ in
segment IV), the dissipated heat $Q_{diss}$ with the energy released
into the cold thermostat $T_C$ in sector II) and assumed
$Q=Q_{in}+Q_{diss}$. In this paper a more refined definition
(Eq. \eqref{eq:input_heat}) of input heat for the case of thermostats with
continuous varying temperatures will be discussed.

If $\langle Q_{in}\rangle>0$ and $\langle W \rangle<0$, efficiency can be
defined as
\begin{equation}
\eta=-\frac{\langle W \rangle}{\langle Q_{in}\rangle}.
\end{equation}


\begin{figure}[!btp]
\centering \includegraphics[width=\columnwidth]{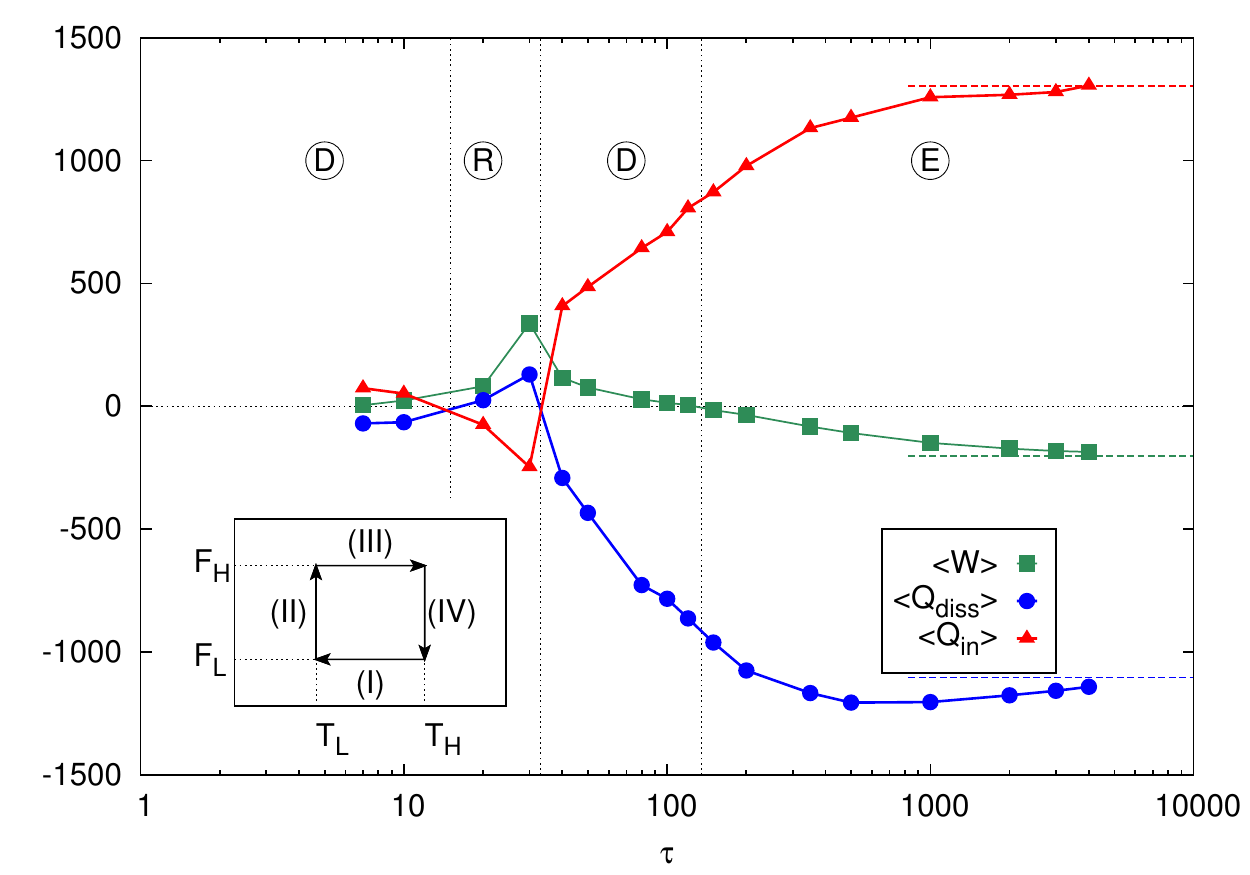}
\caption{Average values per cycle of work $W$ and heats $Q_{diss}$, $Q_{in}$
as a function of the cycle period $\tau$. Dashed horizontal lines represent the adiabatic value of such quantities. Inset: schematic of the cycle protocol in the space of parameters $F,T_o$. Other parameters are $F_L=180$, $F_H=220$, $T_C=11$, $T_H=13$, $m=1$, $M=100$, $N=500$.
 }\label{fig:ave}
\end{figure}

{\em The engine phase diagram} In~\cite{cerino15} we have numerically
studied the MM (the numerical results of the MM reported in this
paper are obtained from numerical simulations based on the common
Verlet scheme with $\Delta t =0.001$), by restricting to a particular
choice of the parameter and changing only the cycle duration
$\tau$. We have seen that in the $\tau \to \infty$ limit the
thermodynamic predictions are recovered, but when $\tau$ is finite the
system behaves differently: in particular there is a stall time
$\tau^*$ where $\langle W \rangle =0$. For $\tau > \tau^*$ the system
produces work as a standard heat engine (``E'' phase). For smaller
$\tau$ the system consumes work in two possible ways: as a
refrigerator (``R'' phase), i.e. by pushing heat from $T_C$ to $T_H$
($\langle Q_{in}\rangle<0$) or as a ``dissipator'' (``D'' phase),
i.e. with the heat going in the natural direction from $T_H$ to $T_C$
($\langle Q_{in}\rangle>0$).  Since the heat extracted from the hot
source, $\langle Q_{in} \rangle$ as a function of $\tau$, crosses
twice the $0$ axis at times $\tau<\tau^*$, there are two changes of
phases (from $D$ to $R$ and from $R$ to $D$). The sequence of phases
is well illustrated in Fig.~\ref{fig:ave}. Interestingly, in the R
phase the absorbed average work per cycle reaches a maximum. In the E
phase the average output power $-\langle W \rangle/\tau$ reaches its
maximum at some value $\tau_{mp}$ where the efficiency appeared to be
close to the CA estimate $\eta_{CA}=1-\sqrt{T_C/T_H}$. A preliminary
study of the fluctuations of $W$, $Q_{in}$ and $\hat{\eta}$ was also
present in~\cite{cerino15}: the statistics of the work or heat per
cycle appeared to be Gaussian in all phases, while the statistics of
the fluctuating efficiency displayed a power law tail at large values,
compatible with a $-3$ exponent.

{\em 2V model} It is reasonable to expect that when the
mass $M$ of the piston is large its dynamics is smoother and possibly
close to a continuous stochastic process. Based upon this idea, in~\cite{cerino15}, apart from an exactly derivable but still complicate model with three variables~\cite{chiuchiu15}, we
proposed a super-simplified model for only two ``slow'' variables:
$X(t)$, $V(t)$:
\begin{equation}\label{eq:init}
\ddot{X} = -k(t) X +\mu(t) \dot{X} + \frac{F(t)}{M} + \sqrt{\frac{2\mu(t) T_o(t)}{2}}\xi(t).
\end{equation}
where ${k(t)=F(t)^2/[M NT_o(t)]}$,
${\mu(t)=2F(t)\sqrt{2m/[ M^2\pi T_o(t)]}}$ and $\xi(t)$ is white
noise with zero average and unitary variance. We assumed the
parameters to vary in a way which is more convenient for calculations
(setting $\omega=2\pi/\tau$ for the engine frequency)
\begin{eqnarray}\label{eq: parametri} 
F(t)&=&F_0(1+\epsilon\cos(\omega t)),\nonumber\\
T(t)&=&T_0(1+\delta\sin(\omega t)).
\end{eqnarray}


In the linear limit $\epsilon\sim \delta \ll 1$, a perturbative
expansions~\cite{cerino15} leads to a formula for the average
trajectory and work performed over a cycle of duration $\tau$,
i.e. of frequency $\omega$:
\begin{align} 
  \langle X(\omega, t)\rangle&=\frac{N T_0}{F_0}+ A(\omega)\{\delta \sin[\omega t + \phi(\omega)]-\epsilon \cos[\omega t + \phi(\omega)] \},\label{anpos}\\
  \langle W (\omega)\rangle&=  F_0 \pi \epsilon A(\omega)\left[
             \epsilon\sin\phi\left(\omega\right) -\delta\cos\phi\left(\omega\right)\right],\label{anwork}
\end{align}
where 
\begin{align}
  A(\omega)&=\frac{NT_0F_0}{\sqrt{8 m T_0F_0^2\omega^2  +\left(F_0^2-\frac{MNT_0}{\pi}\omega^2\right)^2}}\nonumber,\\
\phi(\omega)&=\arctan\left(\frac{2NF_0T_0\omega}{F_0^2-MNT_0\omega^2}\sqrt{\frac{2m}{\pi T_0}}\right).
\end{align}

The 2V model reproduces qualitatively the numerical results of the MM,
but shows quantitative discrepancies. A study of the
time-autocorrelations of the piston position or velocity in the MM,
even at equilibrium, reveals many relaxation
times~\cite{cerino2014fluctuations}: such a study suggests that a
reasonable number of variables is $5$ (for instance some average
properties of the gas, such as average position, velocity and kinetic
energy could be added to the 2V model), but a multivariate stochastic
process with so many variables (and a consequently large number of
parameters) is far from our aim. The question about the linearity of
the model is also open~\cite{sano2015}.

\section{Linear regime}

\label{linear}
\subsection{General results}
Many general results \cite{VandenBroeck2005,Polettini2014,Verley2014}
regarding the performance of finite-time heat engines are obtained
within the framework of irreversible thermodynamics
\cite{deGrootMazur}.  For this reason and also for having a different
insight into the physics of our model, in this Section we recall some
basic notions on the Onsager formulation of out-of-equilibrium
processes in order to fit our molecular model into this
formalism.  

Every irreversible process is characterized by a non-negative entropy
production (we consider the quantities integrated in time on a cycle
period $\tau$) $\Sigma$ which, in turn, can be expressed as the sum of
$n$ products of some thermodynamic (time-integrated) fluxes $J_i$ with
the associated thermodynamic forces $f_i$~\cite{deGrootMazur}:
\begin{equation}\label{eq:decomposition}
\Sigma=\sum_{i=1}^n J_i f_i.
\end{equation}
The entropy production and all the
fluxes are expected to vanish at equilibrium (i.e. when there are no thermodynamic
forces).  Consequently, when the forces are small, the fluxes can be
expressed as linear combinations of the forces,
\begin{equation}
J_i=\sum_{k=1}^n L_{ik} f_k,
\end{equation}
where $L_{ik}$ are the so-called {\em Onsager coefficients}. Thus, in
the linear regime, the entropy production rate is a quadratic form of
the thermodynamic forces,
\begin{equation}
\Sigma= \sum_{i,k=1}^nL_{ik}f_i f_k.
\end{equation}
The matrix of Onsager coefficients ({\it Onsager matrix}), given the
positivity of $\Sigma$, must be positive-semidefinite.

In many physical problems the identification of fluxes and forces is
unambiguous: that is not the case in the problem we are
considering~\cite{proesmans2015b,proesmans2016}. Our model belongs to the class of hamiltonian systems
${\bf x}$ with Hamiltonian $\mathcal{H}({\bf x}(t),F(t))$ and coupled
to a thermostat at the time-dependent temperature $T(t)$. Both $F(t)$
and $T(t)$ are periodic functions of time, with period $\tau$. Since
the thermostatting dynamics satisfies, at every time $t$, the detailed
balance condition with the equilibrium (Gibbs) distribution at
temperature $T(t)$, the total average entropy production (see Appendix
\ref{app:a}) of the system reads
\begin{equation}\label{eq:entropy_definition}
\Sigma(\tau) =-\int_0^\tau \frac{\langle \dot{Q}(t)\rangle}{T(t)}dt\geq 0,
\end{equation}

where $\dot{Q}$ is the rate of heat absorption from the thermostat. To
obtain a decomposition analogous to Eq. \eqref{eq:decomposition}, it is
useful to express the temperature, following \cite{brandner2015}, as
\begin{equation}\label{eq:temp_dec}
T(t)=\frac{T_0(1-\delta^2)}{(1+\delta)-2\delta \gamma(t)},
\end{equation}
where $T_0=(T_H+T_C)/2$ and
\begin{equation}
\delta=\frac{T_H-T_C}{T_H+T_C}.
\end{equation}
The time-dependence of $T(t)$ is expressed through the function
$0\le\gamma(t)\le 1$, so that $\gamma(t)=1 \iff T(t)=T_H=T_0(1+\delta)$ and
$\gamma(t)=0 \iff T(t)=T_C=T_0(1-\delta)$. We
will also use the notation $F_0=(F_H+F_L)/2$ and
\begin{equation}
\epsilon=\frac{F_H-F_L}{F_H+F_L},
\end{equation}
to indicate, respectively, the intermediate force and the relative
excursion. By plugging Eq. \eqref{eq:temp_dec} into
Eq. \eqref{eq:entropy_definition} and using
$\langle W \rangle+\langle Q\rangle =0$ one gets
\begin{equation}
  \Sigma(\tau) = \frac{\langle W \rangle}{T_0(1-\delta)} + \delta \left(\frac{2}{T_0(1-\delta^2)}\int_0^\tau \langle \dot{Q}\rangle \gamma(t) \right).
\end{equation}
By interpreting $\epsilon$ and $\delta$ as the two (adimensional)
independent thermodynamic forces, or affinities, acting on the system,
and recalling that $\langle W \rangle \propto \epsilon$ for small
$\epsilon$, one gets an expression analog to
Eq. \eqref{eq:decomposition}:
\begin{equation}\label{eq:def_entropy_production}
  \Sigma(\tau)= \epsilon  J_1(\tau) +\delta J_2(\tau),
\end{equation}
where
\begin{align}
  J_1(\tau)&=\frac{\langle W \rangle}{\epsilon T_0 (1-\delta)},\label{eq:j1}\\
  J_2(\tau)&=\frac{2}{T_0(1-\delta^2)}\int_0^\tau \langle
  \dot{Q}\rangle \gamma(t)\label{eq:j2}.
\end{align}
The physical meaning of $J_2$ can be understood by analyzing the
limiting case where $T(t)$ is a square wave function between $T_C$ and
$T_H$ (i.e. when $\gamma(t)$ assumes only the values $\gamma=1$ or
$\gamma=0$): in this case $J_2$ is proportional to the heat $Q_H$
exchanged with the hot thermostat.  It is worth noticing that
expressions different from Eq. \eqref{eq:temp_dec} can lead to
different (legit) definition of fluxes: nonetheless, this
decomposition is particularly suited for an analysis of the efficiency
at maximum power (see Sec. \ref{sec:eff_max_pow}).

For small values of $\epsilon$ and $\delta$, i.e. in the linear
regime, the fluxes are linear function of the forces
\begin{equation}
J_i(\tau)=L_{i1}(\tau)\epsilon +L_{i2}(\tau)\delta, \hspace{0.5 cm} i=1,2,
\end{equation}
where the Onsager coefficients $L_{ij}$ non-trivially depend on the
total time $\tau$ of the transformation.
\subsection{Reciprocity relations and behavior far from the linear regime}

We now discuss a generalization of the Onsager reciprocity relations
for systems undergoing cyclical transformations, proposed in in
Ref. \cite{brandner2015} . For each protocol determined by $T(t)$ and
$F(t)$, it is possible to construct its ``time-reversed'' counterpart
$\tilde{T}(t)=T(\tau-t)$ and $\tilde{F}(t)=F(\tau-t)$: if we indicate
with $\tilde{\cdot}$ quantities measured in the time-reversed cycle,
the following relation is a direct consequence of the reversibility of
the microscopic dynamics:
\begin{equation}
L_{12}(\tau)=\tilde{L}_{21}(\tau).
\end{equation}
For the Ericsson protocol described in Sec. \ref{sec:model} the time
reversal transformation can be obtained by taking the same form of the
protocol for $T$ and $F$ with an (inessential) global shift of phase
$t_0=\tau/2$ and inverting the sign of the force difference
$\epsilon\to -\epsilon$. For this reason $\tilde{L}_{21}=-L_{21}$,
i.e.
\begin{equation}
L_{12}(\tau)=-L_{21}(\tau).
\end{equation}
\begin{figure}[!tbp]
\includegraphics[width=\columnwidth]{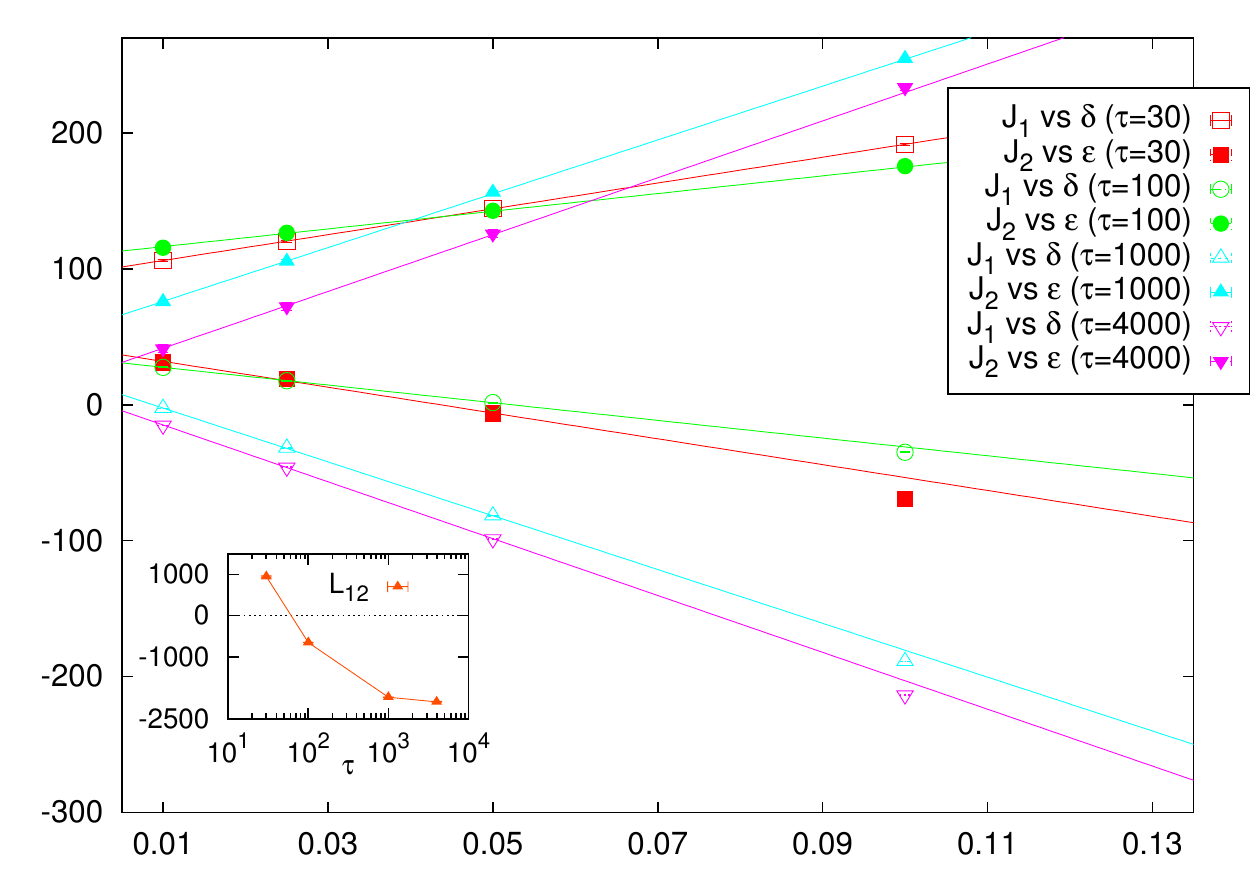}
\caption{Thermodynamic currents $J_1$ (empty symbols) and $J_2$ (solid
  symbols) as functions, respectively, of thermodynamic forces
  $\delta$ and $\epsilon$ for different values of $\tau$. Straight
  lines are obtained by fitting the two data sets at the same $\tau$
  with two linear functions with slope $L_{12}$ and $-L_{12}$
  respectively: for this reason, lines with the same color have
  opposite slope. Inset: Off-diagonal Onsager coefficient $L_{12}$
  obtained with the above procedure as a function of $\tau$. All the
  other parameters are the same of
  Fig. \ref{fig:ave}. }\label{fig:reciprocity}
\end{figure}

\begin{figure}[!htbp]
\includegraphics[width=\columnwidth]{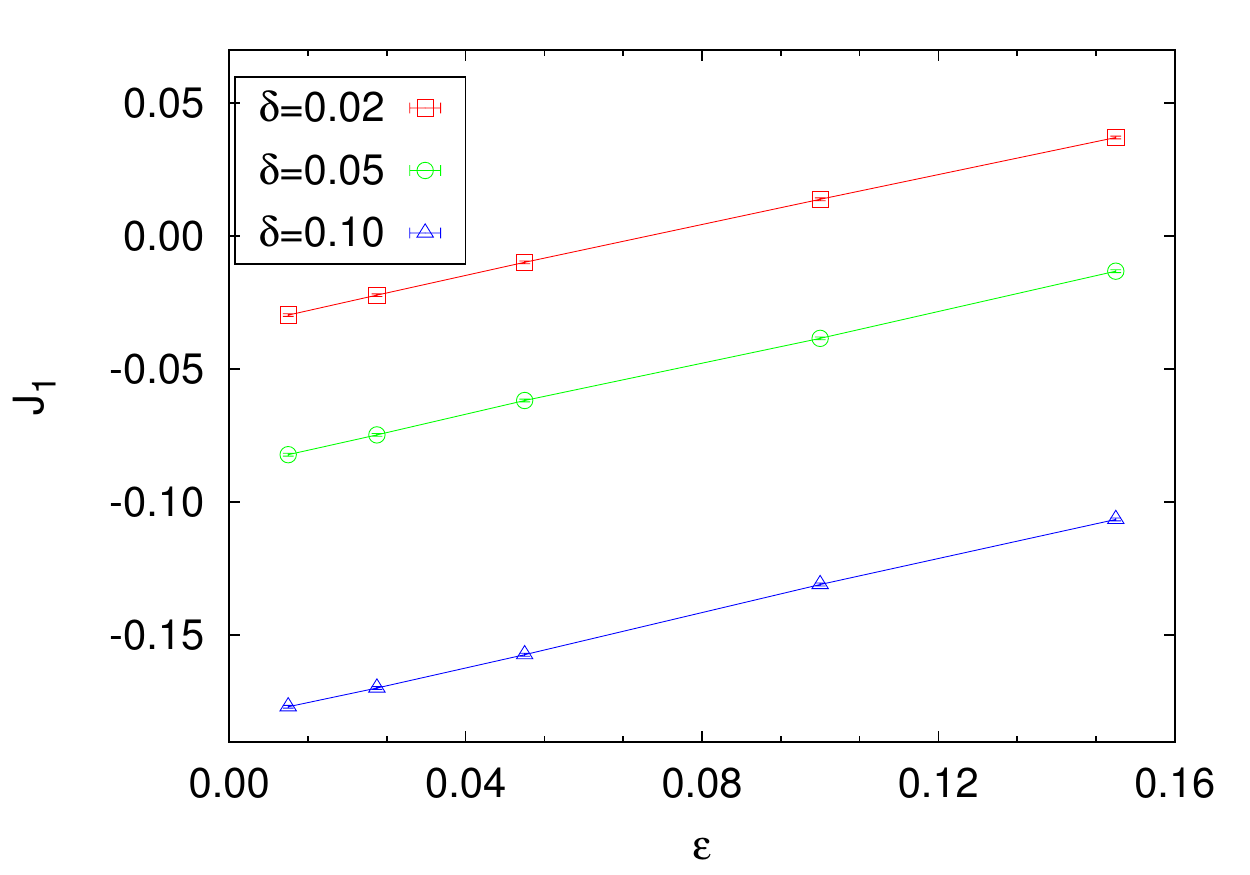}
\caption{The thermodynamic current $J_1$ as a function of $\epsilon$
  for different values of $\delta$ at $\tau=500$. Other parameters are
  the same of Fig.\ref{fig:ave} }\label{fig:parallelism}
\end{figure}

In Fig. \ref{fig:reciprocity} the results of a measurement of $J_1$
and $J_2$ for different values of $\epsilon$ and $\delta$ in molecular
dynamics simulations of the MM are reported. By fixing,
respectively, $\delta=0.05$ or $\epsilon=0.05$, $J_2$ and $J_1$ are
plotted as functions of $\epsilon$ and $\delta$. A linear dependence
is obtained for small values of the thermodynamic forces: moreover,
the data are compatible with the hypothesis of two linear relations
with opposite coefficients describing the functional dependence of
$J_1$ on $\delta$ and of $J_2$ on $\epsilon$ (straight lines in
Fig. \ref{fig:reciprocity}). The measurement also confirms the fact
that the Onsager coefficients have a non trivial dependence on the
total time of the transformation $\tau$ (inset of
Fig. \ref{fig:reciprocity}). In Fig. \ref{fig:parallelism} we report a
measurement of $J_1$ as a function of $\epsilon$ for different values
of $\delta$: since the curves are parallel straight lines, the Onsager
coefficient $L_{11}$ does not depend on the value of $\delta$
(analogous results, not reported here, can be obtained for all the
Onsager coefficients).

\begin{figure}[!btp]
\includegraphics[width=\columnwidth]{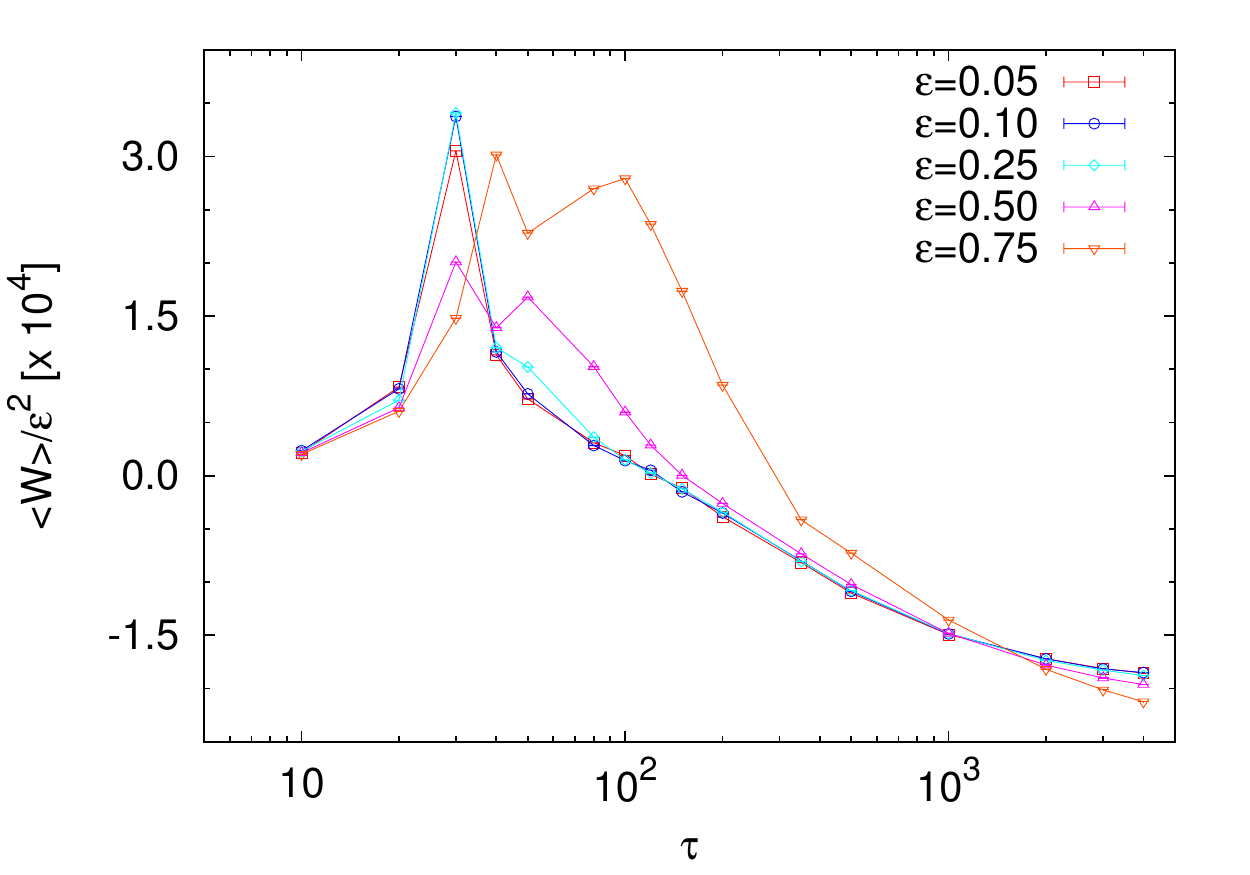}
\caption{The average work per cycle, divided by $\epsilon^2$ is
  reported as a function of $\tau$ for different values of
  $\epsilon$. The other parameters are $\delta=\epsilon$,
$F_0=200$, $T_0=12$, $N=500$, $M=100$, $m=1$.}\label{fig:scaling}
\end{figure}

In Fig. \ref{fig:scaling} we study the limits of the linear behavior
of the MM: by taking $\delta =\epsilon$, we report the average
work divided by $\epsilon^2$ as a function of $\tau$ for different
values of $\epsilon$. In the linear regime the different curves, when
rescaled, must superimpose: this is the case, of course, for small
values of $\epsilon$. At larger values of $\epsilon$ the appearing
discrepancies are not uniform in $\tau$. In particular we remark that
around the maximum of $\langle W\rangle$ the separation is much more
prominent (also signaled by the appearance of a second local maximum
for $\epsilon\ge 0.25$). Note that the non-linearity appears also in
the large $\tau$ limit, since higher order terms of the expansion of
the quantity $\log\left(\frac{1+\epsilon}{1-\epsilon}\right)$ in the
adiabatic formula of work (see Table \ref{tab: 1}) become
relevant. Unfortunately an analytic description of the non-linear
regime is not yet available and the interesting features of such a
regime will hopefully be the subject of future investigations.

\subsection{Analytic expression of the Onsager coefficients in the 2V  model}
\label{sec:onsager_2v}

In the simplified 2V model, Eq. \eqref{eq:init}, it is possible to
obtain an explicit expression for the above mentioned Onsager
coefficients $L_{ij}$.  By plugging Eq. \eqref{anwork} into definition
\eqref{eq:j2} one immediately gets the linear expansion for $J_1$,
i.e.
\begin{equation}\label{eq:j1_explicit}
  J_1(\tau)=\frac{F_0}{T_0}\pi A\left(\frac{2\pi}{\tau}\right)\left(\epsilon \sin \phi\left(\frac{2\pi}{\tau}\right) -\delta \cos \phi\left(\frac{2\pi}{\tau}\right)\right).
\end{equation}
In order to get the corresponding expansion for $J_2$, we start by
plugging the protocol $T(t)=T_0(1+\delta \sin(2\pi t/\tau))$,
Eq. \eqref{eq: parametri}, into Eq. \eqref{eq:temp_dec} and get
\begin{equation}
  \gamma(t)=\frac{1}{2}\left(1 + \sin\left(\frac{2 \pi}{\tau} t\right)\right) + \mathcal{O}(\delta).
\end{equation}
To obtain an explicit expression for $\langle \dot{Q}\rangle$ it is
necessary to substitute the asymptotic solution for $\langle
X\rangle$, Eq. \eqref{anpos} into the expression for energy $\langle
E(t)\rangle= N T_o(t)/2 +M \langle \dot{X}\rangle^2/2 +F \langle
X\rangle$, and then use the definition of heat $\langle \dot{Q}\rangle
=\langle \dot{E}\rangle -\dot{F}\langle
X\rangle=\mathcal{O}(\epsilon,\delta)$.  Retaining only first order
terms in $\epsilon$ and $\delta$ of Eq. \eqref{eq:j2} gives the
following expression for $J_2$:
\begin{equation}\label{eq:j2_explicit}
  J_2=\frac{\pi F_0}{T_0}  A\left(\frac{2\pi}{\tau}\right) \left(\epsilon \cos\phi\left(\frac{2 \pi}{\tau}\right)+\delta \sin\phi\left(\frac{2\pi}{\tau}\right)\right).
\end{equation}
In summary, the Onsager matrix takes the form
\begin{equation} \label{onsag} L = \frac{F_0}{T_0}\pi
  A\left(\frac{2\pi}{\tau}\right) \begin{pmatrix} \sin
    \phi\left(\frac{2\pi}{\tau}\right) &-\cos
    \phi\left(\frac{2\pi}{\tau}\right) \\ \cos
    \phi\left(\frac{2\pi}{\tau}\right) &\sin
    \phi\left(\frac{2\pi}{\tau}\right)\end{pmatrix}
\end{equation}
The anti-reciprocal relation $L_{12}=-L_{21}$ in the Onsager matrix is
due to the fact that the protocol used in the 2V model (Eq. \eqref{eq:
  parametri}) behaves, under time-reversal, in the same way as the
Ericsson protocol. We can also get a very simple expression for the
total entropy production,
\begin{equation}\label{eq: entropy_production}
  \Sigma(\tau)=\frac{ \pi F_0}{T_0} A\left(\frac{2\pi}{\tau}\right) \sin\phi\left(\frac{2\pi}{\tau}\right)(\epsilon^2+\delta^2),
\end{equation}
which, as expected, is always positive because $A(\omega)\geq 0 $ and
$0\leq \phi(\omega)\leq \pi$.

\section{Efficiency at maximum power: linear regime and beyond}
\label{sec:eff_max_pow}

The aim of this Section is to study the efficiency at maximum power of
our engine comparing three different levels of approximation: numerical simulations of the full Molecular Model, numerical
solutions of its coarse-grained 2V version, and analytical solutions
of the 2V model for small values of $\epsilon$ and $\delta$ (i.e. when
fluxes are linear in the forces).

First, we need to find a suitable definition of efficiency $ \eta$ for
our case. Usually the efficiency of a heat engine working at contact
with two thermostats (at temperatures $T_C<T_H$) is simply the ratio of
the output work divided by the energy absorbed from the hotter
thermostat $Q_{in}$. For transformations that involve thermostats at
temperatures ranging continuously in the interval $T_C<T(t)<T_H$, the
input heat must be redefined \cite{brandner2015} as
\begin{equation}\label{eq:input_heat}
Q_{in}=\int_0^\tau \dot{Q}(t) \gamma(t) dt,
\end{equation}
where $\tau$ is the total time of the transformation and $\gamma(t)$
is the function appearing at the denominator of the right-hand side of
Eq. \eqref{eq:temp_dec}. This definition comes from the observation,
already reported in the previous section, that
Eq. \eqref{eq:input_heat} gives the correct result in the simple case
of two thermostats at $T_C$ and $T_H$: moreover for the Ericsson
protocol this expression reduces, in the quasi-static limit, to the
heat extracted from the hot reservoir $T_H$.

From this definition, we get the expression for the efficiency of the engine,
\begin{equation}\label{eq:efficiency_def}
  \eta=\frac{-\langle W \rangle}{\langle Q_{in}\rangle}=-\frac{2\epsilon J_1}{(1+\delta) J_2},
\end{equation}
where we used Eqs. \eqref{eq:j1} and \eqref{eq:j2} to recast the
expression in terms of the thermodynamic currents $J_1$ and $J_2$.
When the total average entropy production $\Sigma$,
Eq. \eqref{eq:def_entropy_production}, vanishes, it is straightforward
to prove that the efficiency assumes the Carnot value
$\eta_C=1-T_c/T_H=2\delta/(1+\delta)$. In the 2V model, the entropy
Eq.  \eqref{eq: entropy_production}, can vanish only for $\phi=0$,
i.e. in the adiabatic limit $\tau\to \infty$ (or in the trivial,
non-interesting case $\tau=0$). It is quite reasonable to assume that,
also in the general case, a vanishing entropy can be only obtained by
varying very slowly the external parameters. As a consequence, the
power corresponding to a maximally efficient engine must be zero. For
this reason, in order to characterize the performance of the engine,
we will study the efficiency at maximum power (EMP) $\tilde{\eta}$,
i.e the efficiency corresponding to a choice of the external
parameters that maximizes the output power.  In the last decades, a
series of important results were obtained regarding the EMP: perhaps
the most notable is that, under some rather general assumptions
\cite{curzon1975efficiency,VandenBroeck2005}, a universal bound for
the EMP is given by the so-called Curzon-Ahlborn efficiency
$\eta_{CA}$,
\begin{equation}\label{eq:curzon-ahlborn}
\tilde{\eta}\leq \eta_{CA}=1-\sqrt{\frac{T_C}{T_H}}\approx \frac{1}{2}\eta_C + \mathcal{O}(\eta_C^2).
\end{equation}
We now investigate the validity of such a bound, and compare our
results with other recent works. 

In our models the output power:
\begin{equation}\label{eq:output_power}
  \mathcal{P}=-\frac{\langle W \rangle}{\tau}
\end{equation}
depends on the details of the model ($N,M,m$) as well as on the choice
of the external protocol ($T_0,F_0,\epsilon,\delta,\tau$). Since the
engine working state consists in a perturbation of the equilibrium
state determined by $T_0$ and $F_0$, it appears quite natural, in
order to maximize $\mathcal{P}$, to fix $N$, $m$, $M$, $T_0$ and
$F_0$.  Moreover, since $\eta_C$ and $\eta_{CA}$ only depend on the
value of $\delta$, we also fix the temperature difference and
therefore maximize the two-variables function
$\mathcal{P}(\epsilon,\tau)$.

\subsection{Linear regime}

\begin{figure}[!hbtp]
\includegraphics[width=\columnwidth,clip=true]{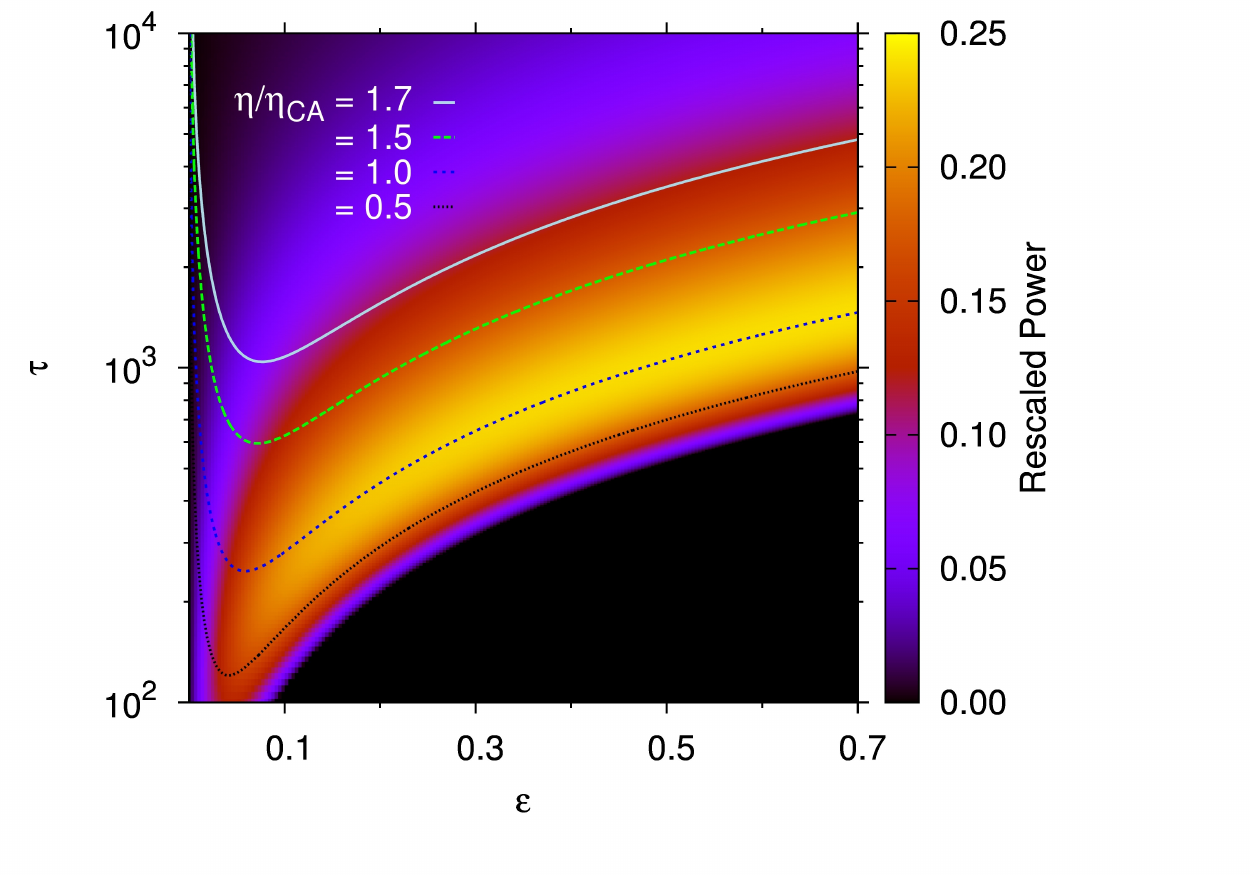}
\caption{{\em 2V model, linear appoximation}: contour plot of rescaled $\mathcal{P}(\epsilon,\tau)$ as a function of $\tau$ and $\epsilon$. The
  power is rescaled by the factor $w_{ad}^{2V}=\epsilon \delta N T_0
  \pi/100$ evaluated in $\epsilon=0.1$, $\delta=1/12$, $N=500$,
  $T_0=12$, $w_{ad}^{2V}\approx 1.57$. Continuous line represents the
  set of points $(\epsilon,\tau)$ where $\eta/\eta_{CA}$ is
  constant.}\label{fig:contourplot}
\end{figure}

In Fig. \ref{fig:contourplot} the linear approximation for the output
power in the 2V model (obtained by plugging Eq. \eqref{anwork} into
Eq. \eqref{eq:output_power}) is plotted as a function of $\epsilon$
and $\tau$. In view of a comparison between the results in the 2V
model and in the molecular model, whose protocols are slightly
different, we rescaled the work and the power so that, for
$\epsilon=0.1$ the asymptotic value for work in the limit
$\tau\to\infty$ is fixed to $\lim_{\tau \to
  \infty}W_{rescaled}(\epsilon=0.1,\tau)=100$. The plot shows that (in
this linear approximation) the power does not reach a global maximum
at a unique value of $(\tau,\epsilon)$. Indeed, there exists a curve
$\tau_{mp}(\epsilon)$ consisting of $\tau$-maxima points, i.e. where
$\partial_\tau \mathcal{P}=0$.  The maximum power curve
$\mathcal{P}[\epsilon,\tau_{mp}(\epsilon)]$ saturates to a constant
value for increasing $\epsilon$. In addition we also note that
$\tau_{mp}(\epsilon)$ is an increasing function of $\epsilon$,
eventually saturating at the value $\lim_{\epsilon\to
  1}\tau_{mp}\approx 1500$. In the plot we have also shown the curves
at constant $\eta$: it is interesting to notice that the
$\tau_{mp}(\epsilon)$ approaches the curve where $\eta=\eta_{CA}$ at increasing $\epsilon$.

\begin{figure}[!btp]
\includegraphics[width=\columnwidth]{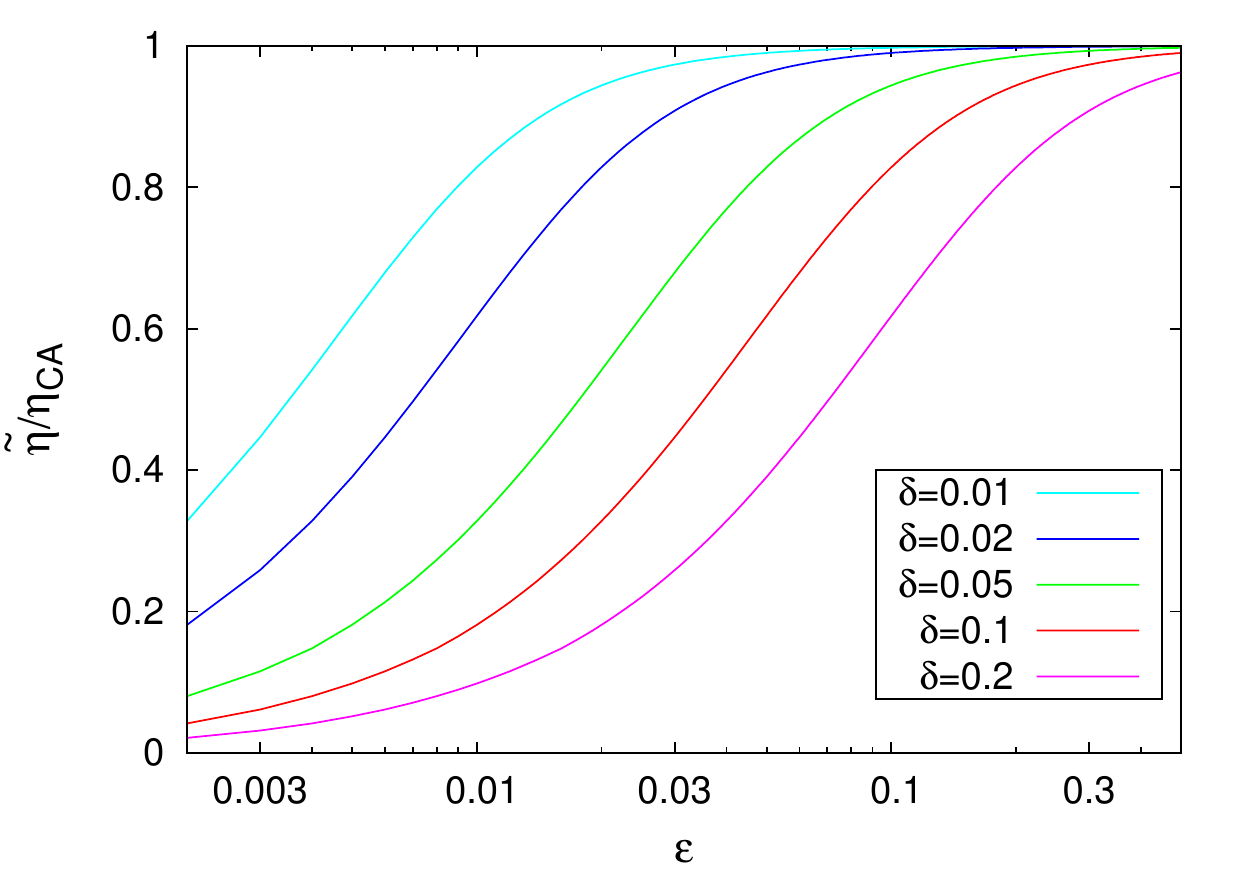}
\caption{{\em 2V model, linear approximation}: ratio of efficiency at
  maximum (with respect to $\tau$) power $\tilde{\eta}$ and
  Curzon-Ahlborn efficiency $\eta_{CA}$, as a function of $\epsilon$
  for different values of $\delta$. Other parameters are $N=500$,
  $M=100$, $m=1$, $T_0=12$, $F_0=200$.}\label{fig:eff_max_pow_2v}
\end{figure}

The EMP curves, $\tilde{\eta}(\epsilon,\tau_{mp}(\epsilon))$ for
different values of $\delta$ are shown in
Fig. \ref{fig:eff_max_pow_2v}: we observe that - consistently with the
previous observation - the CA efficiency is only reached for large
values of $\epsilon$ where, in principle, the linear approximation is
no more reliable. However, by decreasing $\delta$ a faster convergence
toward the CA efficiency is observed: this suggests the possibility to
observe $\tilde{\eta}=\eta_{CA}$ even in the linear regime.

Let us remark again that in our system it is possible to separate the
time $\tau$ of the transformation from the small force limit (small
$\epsilon$ and $\delta$): this means that we are able to consider a
linear approximation (and construct the corresponding $\tau$-dependent
Onsager matrix) which is valid, in the small $\epsilon-\delta$ limit,
at every value of $\tau$.  On the contrary, in many recent papers (see
e.g. \cite{Izumida2009,proesmans2015}) one of the small thermodynamic
forces must be the inverse of the time of the transformation $\tau$.

\subsection{Non-linear regime}

\begin{figure}[!tp]
\includegraphics[width=\columnwidth]{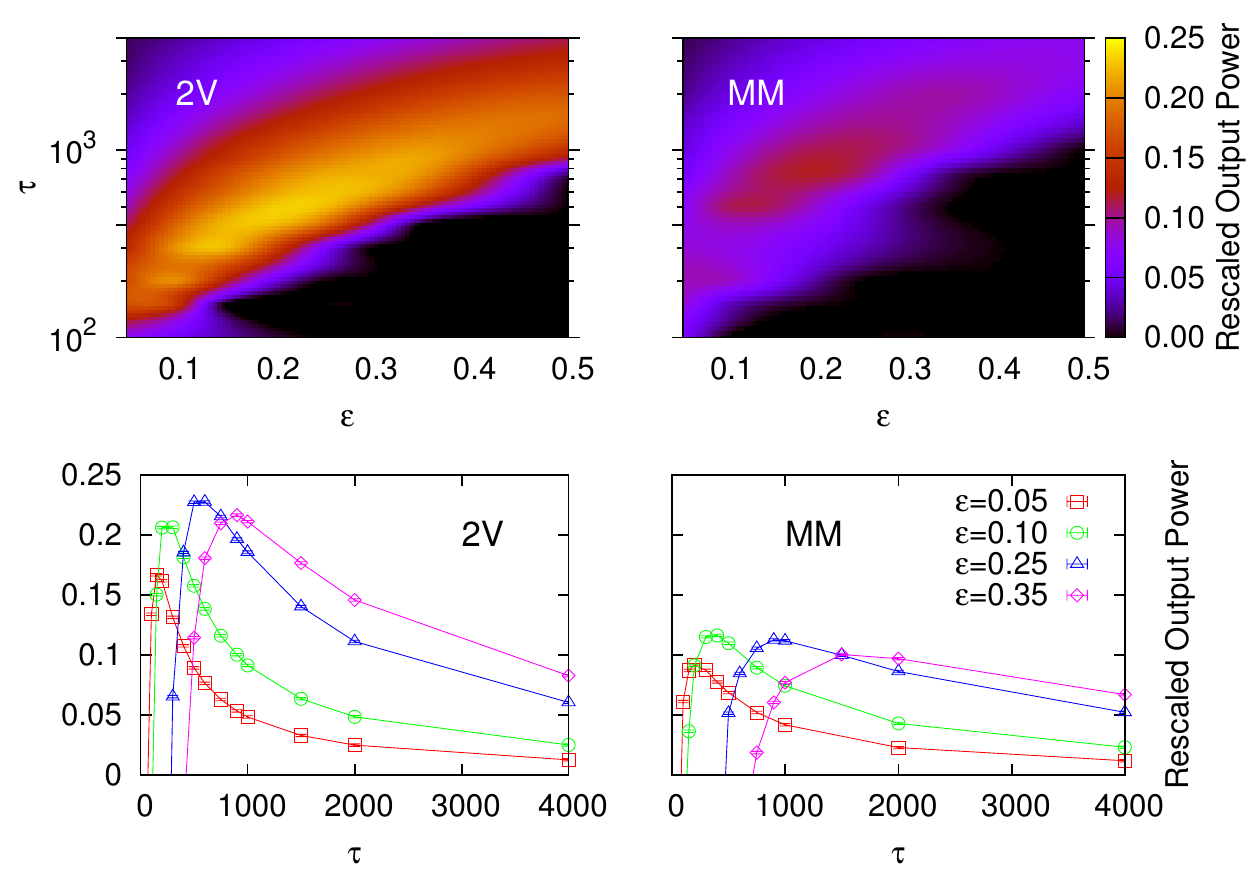}
\caption{{\em Left: 2V model solved numerically, right: simulations of
    the MM.} Top panels: colour plot of the rescaled output power. Bottom panels: the rescaled output power as a function of $\tau$
  for different values of $\epsilon$.  Parameters are the same in 2V
  and MM simulations: $\delta=1/12$, $N=500$, $M=100$, $m=1$,
  $F_0=200$, $T_0=12$. The power in MM simulation is divided by the
  quantity $w_{ad}^{MM}=0.02\, N\, T_0\,\delta
  \ln\left((1+\epsilon)/(1-\epsilon)\right)$ in $\epsilon=0.1$,
  $w_{ad}^{MM}\approx 2.007$. In the 2V model the rescaling factor is
  $w_{ad}^{2V}= \epsilon \delta N T_0 \pi/100$ which, in
  $\epsilon=0.1$ gives $w_{ad}^{2V}\approx
  1.57$}\label{fig:simulazioni}
\end{figure}

The absence of an absolute maximum for the power, however, appears
only to be a consequence of the linear approximation used to solve the
2V equations. Indeed, by performing numerical integration of the full
2V model (Eq. \eqref{eq:init}) and simulations of the MM, we observe a
rather different situation, which is reported in
Fig. \ref{fig:simulazioni}: the two top panels represent the color map
of the functions $\mathcal{P}(\epsilon,\tau)$ for the two models, the
two bottom panels show some sections $\mathcal{P}(\epsilon^*,\tau)$ vs
$\tau$, for some values of $\epsilon^*$. By analyzing these last
plots, we observe that the maximum power increases when going from
$\epsilon=0.1$ to $\epsilon=0.25$ and then decreases again in
$\epsilon=0.35$. This suggests that is indeed possible, at least
numerically, to find a specific value for $\epsilon$ and $\tau$
corresponding to the global maximum power. The only significant
difference between the 2V model and the MM is that the output power is
smaller, in general, than the one obtained in the 2V model.

\begin{figure}[!tp]
\includegraphics[width=\columnwidth]{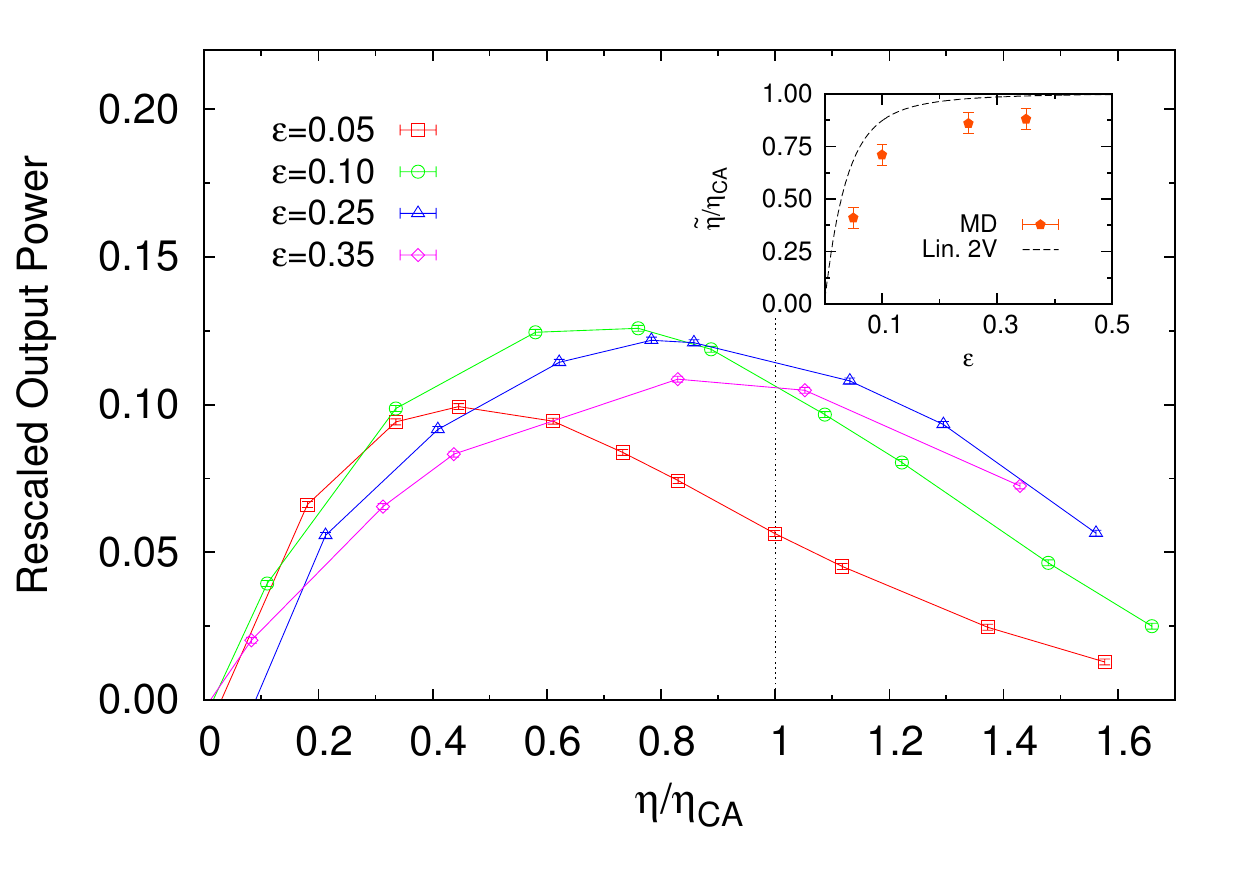}
\caption{{\em Simulations of the MM}: the rescaled output power is
  reported as a function of the rescaled efficiency $\eta/\eta_{CA}$
  for different values of $\epsilon$. The rescaling factor for power
  is $w_{ad}^{MM}\approx 2.007$ and
  $\eta_{CA}=1-\sqrt{(1+\delta)/(1-\delta)}$. Inset: comparison
  between numerically solved 2V model and simulations of the MM for the
  relative EMP $\tilde{\eta}/\eta_{CA}$ is reported as a function of
  $\epsilon$. Other parameters
  are the same as in Fig. \ref{fig:simulazioni}. }\label{fig:emp_MD}
\end{figure}

In Fig. \ref{fig:emp_MD} we focus on the MM and report the same output
power as a function of the efficiency (which, interestingly, is a
bijective function of $\tau$). We observe that, at every value of
$\epsilon$, the maximum power is attained at a value $\eta
<\eta_{CA}$: moreover the global maximum power corresponds to an
efficiency that is approximately the 70\% of the Curzon-Ahlborn
efficiency (i.e. 35\% of the Carnot efficiency). The Curzon-Ahlborn
efficiency seems to be approached for larger values of $\epsilon$. A
comparison between the EMP measured in the numerical simulations of
the MM  and the corresponding result in the 2V model (inset of
Fig. \ref{fig:emp_MD}), shows that the simplified model overestimates
the actual value of $\tilde{\eta}$.

We wish to spend a few words about the observed lower value of the EMP with respect to the CA
efficiency. We see that it is a consequence of the widening of the space of
parameters~\cite{bauer16}. In Ref. \cite{VandenBroeck2005} (and its generalization to
non-symmetric Onsager matrices in Ref. \cite{brandner2015}), it is
proved that, for fixed Onsager coefficients (i.e. for fixed $\tau$)
$\eta_{CA}$ is reached whenever the value
\begin{equation}
q=\frac{L_{12}L_{21}}{L_{11}L_{22}-L_{12}L_{21}}
\end{equation}
is close to $q=1$ (``tight coupling'' hypothesis). When the Onsager
matrix is $\tau$-dependent, the variable $q$ is a function of $\tau$,
$q=q(\tau)$. Suppose the existence of a value of $\tau=\tau_0$ such
that the tight-coupling hypotesis is verified $q(\tau_0)\simeq 1$:
then, by denoting with $\epsilon_0$ the value of $\epsilon$ that
maximizes the power at $\tau=\tau_0$ (with respect to $\epsilon$), we
will obtain $\eta(\tau_0,\epsilon_0)=\eta_{CA}$.  On the other hand,
the global maximum power in the $(\tau,\epsilon)$-plane may occur in a
point $(\tau_1,\epsilon_1)$ for which the tight coupling condition is
violated $q(\tau_1)<1$, corresponding to an efficiency
$\eta(\tau_1,\epsilon_1)<\eta_{CA}$.  To summarize, extending the
space of parameters, e.g. by allowing $\tau$ to vary, may permit to
find a larger maximum power, but this does not guarantee that the
corresponding efficiency would be closer to the CA efficiency.


\section{Conclusions}
\label{concl}

We have studied the thermodynamic properties of a model engine. The
essential distinguishing features of our system are: 1) a realistic
gas-like dynamics occurring in a spatially extended domain (i.e. the
space between the moving piston and the thermostat); 2) inertial
effects which allow for a larger freedom in the choice of parameters
(e.g. $\tau$ is not constrained by the piston's velocity) and a more
rich phase diagram; 3) a cyclical protocol repeating in a finite time
$\tau$ which is not related to the relative excursions of the pressure
and temperature, $\delta$ and $\epsilon$. The results of the
simulations of the molecular model are compared to analytical and
numerical solutions of a simplified, coarse-grained, equation, which
yields a qualitatively similar picture. A clear scenario emerges from
our study, where the relation between the fluxes (heat and work) and
the thermodynamic forces do not depend trivially upon $\tau$, as
it appears, for instance, in the approximated expressions of the
Onsager matrix, Eq.~\eqref{onsag}. Our model is appropriate to study
the issues of finite-time thermodynamics in a case where the adiabatic
limit ($\tau \to \infty$) and the linear regime (small thermodynamic
forces) are disentangled. It would be interesting to check whether
higher order terms in the expansion (in $\delta,\epsilon$) of the 2V
model is able to reproduce the presence (observed both in the
non-linear 2V model and the MM ) of a global maximum of the
power. An interesting future extension of the present study is taking
into account more realistic molecular interactions. A promising line
of investigation, in view of the finite number $N$ of particles in the
engine, concerns the study of fluctuations of heat and work and the
consequent definition of a fluctuating efficiency~\cite{Verley2014},
already partially discussed in~\cite{cerino15}.

\appendix

\section{}
\label{app:a}
For the sake of consistency, in this Appendix we prove
Eq. \eqref{eq:entropy_definition} for the simple case of discrete
phase space and time. This formula, which holds also if time and space
are both continuous, has appeared many times in the literature (see
e.g.  Ref.~\cite{Gong2015} for a nice pedagogical derivation).

Let us consider a discrete Markov process with time-dependent
transition probabilities $p(i\to j,t)$ satisfying the detailed balance
(DB) condition
\begin{equation}
e^{-\frac{\mathcal{H}(i,\lambda_t)}{T_t}}p(i\to j,t)=e^{-\frac{\mathcal{H}(j,\lambda_t)}{T_t}}p(j\to i,t),
\end{equation}
for every value of $i,j$ and $t$, where $\lambda_t$ is an external
time-dependent protocol and $T_t$ the time-dependent temperature. The
entropy production of the medium for a given trajectory
$\{i_t\}_{t=0}^\tau$ reads \cite{Lebowitz1999}
\begin{equation}\label{eq:app_entropy_prod}
\Sigma_{m}(\tau)=\log \frac{P[\{i_s\}_{s=0}^\tau|i_0]}{\tilde{P}[\{i_{\tau-s}\}_{s=0}^\tau|i_\tau]}
\end{equation}
where $\tilde{P}$ denotes the probability of the trajectory in
a process with the time reversed protocol $\tilde{\lambda}_t=\lambda_{\tau-t-1}$.
By expliciting Eq. \eqref{eq:app_entropy_prod} and using DB one gets
\begin{align}
  \Sigma_m(\tau)&=\log \frac{p(i_o\to i_1,\lambda_0)\ldots p(i_{\tau-1}\to i_\tau,\lambda_{\tau -1})}{p(i_\tau\to i_{\tau -1},\lambda_{\tau-1})\ldots p(i_1\to i_0,\lambda_0)}=\nonumber\\
&=-\sum_{t=1}^{\tau} \frac{\mathcal{H}(i_t,\lambda_{t-1})-\mathcal{H}(i_{t-1},\lambda_{t-1})}{T_{t-1}}.
\end{align}
At every step the energy difference of the system can be decomposed according to
\begin{equation}
\Delta E= \mathcal{H}(i_t,\lambda_t)-\mathcal{H}(i_{t-1},\lambda_{t-1})=W_t+Q_t,
\end{equation}
where the work $W_t$ is the contribution due only to the change of
$\lambda$,
$W_t=\mathcal{H}(i_t,\lambda_t)-\mathcal{H}(i_t,\lambda_{t-1})$, and
the heat $Q_t$ is due to the change of state at fixed $\lambda$,
$Q_t=\mathcal{H}(i_t,\lambda_{t-1})-\mathcal{H}(i_{t-1},\lambda_{t-1})$. Consequently
\begin{equation}
\Sigma_m(\tau)=-\sum_{t=0}^{\tau-1} \frac{Q_t}{T_t}.
\end{equation}
The non-negative total entropy production $\Sigma_{tot}(\tau)$
\cite{Seifert2005} is the sum of the medium entropy production
$\Sigma_m$ and the system entropy production
\begin{equation}
\Sigma_s(\tau)=-\log\frac{\rho(i_0,t=0)}{\rho(i_\tau,t=\tau)},
\end{equation}
where $\rho(i_t,t)$ is the probability distribution function (pdf) of
the system at time $t$. If $\lambda_t$ and $T_t$ are periodic function
of time with period $\tau$, there exists a periodic ``stationary'' pdf
$\rho_\infty$ such that $\rho_\infty(i,t)=\rho_\infty(i,t+\tau)$, for
every $i$ and $t$. This also means that if the initial pdf
$\rho(i,t=0)=\rho_\infty(i,t=0)$, on average, the system entropy
production vanishes, $\langle \Sigma_s\rangle=0$ and finally
\begin{equation}
\langle \Sigma(\tau)\rangle=\langle \Sigma_m \rangle=-\sum_{t=0}^\tau \frac{\langle Q \rangle}{T_t},
\end{equation}
which is the discrete time equivalent of Eq. \eqref{eq:entropy_definition}. 
 \bibliography{mergedbiblio}

\end{document}